\documentclass[12pt,a4paper,final]{iopart}
\usepackage{iopams}
\usepackage{float}
\usepackage{graphicx}
\UseRawInputEncoding   
\usepackage[breaklinks=true,colorlinks=true,linkcolor=blue,urlcolor=blue,citecolor=blue]{hyperref}
\usepackage{tabularx}
\makeatletter

\newcommand{\Rmnum}[1]{\expandafter\@slowromancap\romannumeral #1@}

\makeatother

\begin{document}

\title{La$_2$Rh$_{3+\delta}$Sb$_4$: A new ternary superconducting rhodium-antimonide}

\author{Kangqiao Cheng$^{1}\dag$, Wei Xie$^1$\footnote[1]{These authors contribute equally to this work}, Shuo Zou$^1$, Huanpeng Bu$^2$, Jin-Ke Bao$^{3}$, Zengwei Zhu$^{1}$, Hanjie Guo$^{2*}$, Chao Cao$^{4*}$, and Yongkang Luo$^{1*}$}

\address{$^{1}$ Wuhan National High Magnetic Field Center and School of Physics, Huazhong University of Science and Technology, Wuhan 430074, China}
\address{$^{2}$ Neutron Science Platform, Songshan Lake Materials Laboratory, Dongguan, Guangdong 523808, China}
\address{$^{3}$ Department of Physics, Materials Genome Institute and International Center for Quantum and Molecular Structures, Shanghai University, Shanghai 200444, China}
\address{$^{4}$ Center for Correlated Matter and School of Physics, Zhejiang University, Hangzhou 310058, China}

\ead{hjguo@sslab.org.cn; ccao@zju.edu.cn; mpzslyk@gmail.com}

\date{\today}

\begin{abstract}
Rhodium-containing compounds offer a fertile playground to explore novel materials with superconductivity and other fantastic electronic correlation effects. A new ternary rhodium-antimonide La$_2$Rh$_{3+\delta}$Sb$_4$ ($\delta\approx 1/8$) has been synthesized by a Bi-flux method. It crystallizes in the orthorhombic Pr$_2$Ir$_3$Sb$_4$-like structure, with the space group $Pnma$ (No. 62).
The crystalline structure appears as stacking the two dimensional RhSb$_4$- and RhSb$_5$-polyhedra networks along $\mathbf{b}$ axis, and the La atoms embed in the cavities of these networks. Band structure calculations confirm it as a multi-band metal with a van-Hove singularity like feature at the Fermi level, whose density of states are mainly of Rh-4$d$ and Sb-5$p$ characters. The calculations also imply that the redundant Rh acts as charge dopant. Superconductivity is observed in this material with onset transition at $T_c^{on}\approx$ 0.8 K. Ultra-low temperature magnetic susceptibility and specific heat measurements suggest that it is an $s$-wave type-II superconductor. Our work may also imply that the broad $Ln_2$$Tm_{3+\delta}$Sb$_4$ ($Ln$=Rare earth, $Tm$=Rh, Ir¡­) family may host new material bases where new superconductors, quantum magnetism and other electronic correlation effects could be found. \\

\hspace{-15pt}\textbf{Keywords:} Superconductivity, Electronic correlation effect, Crystalline structure

\end{abstract}


\maketitle

\hspace{-30pt}
\fbox{%
\parbox{\textwidth}{%
\textbf{Future Perspectives}\\
\emph{}\\
The moderate on-site Coulomb repulsion $U$ and spin-orbit coupling strength of Rh-$4d$ electrons endow them with variable, intriguing natures, and this offers a rich platform for materials design. We hereby report a new ternary rhodium-containing compound La$_2$Rh$_{3+\delta}$Sb$_4$ ($\delta\approx 1/8$) that becomes a type-II superconductor at low temperature. More importantly, our work also suggests that the broad family of $Ln_2$$Tm_{3+\delta}$Sb$_4$ ($Ln$=Rare earth, $Tm$=transition metals) - whose physical properties are not yet well explored - likely host new quantum material bases where novel superconductivity, quantum magnetism, quantum criticality and other electronic correlation effects can be found.
}%
}

\section{Introduction}

Ternary rhodium-containing compounds have spanned fertile material bases for studies of unconventional superconductivity (SC) and electronic correlation effects. On the one hand, Rh-4$d$ electrons possess moderate on-site Coulomb repulsion $U$ and spin-orbit coupling (SOC) strength, which endow them with variable - either localized or itinerant - electronic states (e.g., see Refs.~\cite{Perry-Sr2RhO4,LuoY-Li2RhO3}). On the other hand, the orbital configuration $4d^85s^1$ of Rh may lead to a large density of states in the presence of specific crystalline electric field, which enhances inter-site hybridization and in turn facilitates rich physical properties such as heavy-fermion SC observed in CeRh$_2$Si$_2$\cite{Movshovich-CeRh2Si2SC}, CeRhIn$_5$\cite{Hegger-CeRhIn5}, PuRhGa$_5$\cite{Wastin-PuRhGa5SC}, URhGe\cite{Aoki-URhGeSC} and CeRh$_2$As$_2$\cite{Khim-CeRh2As2,Kibune-CeRh2As2NQR}, Kondo insulator in CeRh(As,Sb)\cite{Kumigashira-CeRhAsSbTI}, unconventional quantum criticality in YbRh$_2$Si$_2$\cite{Gegenwart-YbRh2Si2QCP} and CeRh$_6$Ge$_4$\cite{Shen-CeRh6Ge4FMQCP}, and quadrupolar Kondo effect in PrRh$_2$Zn$_{20}$\cite{Onimaru-PrRh2Zn20AFQSC}.


The chemical formula $LnTm_{2-x}$Pn$_2$ ($Ln$=Rare earth, $Tm$=Transition metal, $Pn$=P,As,Sb) stands for a big class of ternary pnictides that attracted a broad audience in recent years. Amongst them, the most well-known phases should be the so-called ``122" family ($x=0$) that consists of two branches of tetragonal crystalline structures: the ThCr$_2$Si$_2$ type ($I4/mmm$, No. 139) and CaBe$_2$Ge$_2$ type ($P4/nmm$, No. 129). Besides these stoichiometric 122 phases, substantial deficient can exist on the transition metal site\cite{Hofmann-1988,Cava-1993,Yang-2005,Luo-CeNi2As2}. It is interesting to note that in some cases, such deficient can maintain the 122-type structure, e.g. CeNi$_{1.75}$As$_2$\cite{Luo-CeNi2As2Pre}, while in other cases, the structure collapses and changes into a new structure. A representative example is Pr$_2$Ir$_3$Sb$_4$ ($x=1$) which belongs to the orthorhombic $Pnma$ space group (No. 62). Multiple compounds have been found to crystallize in this 234 phase, such as $Ln_2$Ir$_3$Sb$_4$ ($Ln$=La, Ce, Pr, Nd)\cite{Schafer-Ce234} and Ce$_2$Rh$_3$Sb$_4$\cite{GilR-Ce234}. In particular, previous studies on the rhodium-antimonide Ce$_2$Rh$_3$Sb$_4$ based on polycrystal samples manifested that the local moments of Ce$^{3+}$ remain paramagnetic down to 2 K\cite{GilR-Ce234}, implying that it probably sits nearby a quantum critical point. Single crystals with high quality are needed to better understand its ground state.

With the goal of growing the single crystalline Ce$_2$Rh$_3$Sb$_4$ and its non-$4f$ reference, we discovered a new ternary rhodium-antimonide La$_2$Rh$_{3+\delta}$Sb$_4$ that crystallizes in the orthorhombic Pr$_2$Ir$_3$Sb$_4$-like structure with redundant Rh. Herein, we report the synthesis of La$_2$Rh$_{3+\delta}$Sb$_4$ single crystals, determination of its crystalline structure, and systematic study of its physical properties by a combination of electrical transport / magnetic / thermodynamic measurements and first-principles band structure calculations. Our work reveals that La$_2$Rh$_{3+\delta}$Sb$_4$ is a new superconducting material with onset temperature $T_c^{on}\approx 0.8$ K, and its normal state is a multi-band metal with van-Hove singularity like feature at the Fermi level.

\begin{figure}[!ht]
\centering
\vspace*{0pt}
\hspace*{0pt}
\includegraphics[width=14cm]{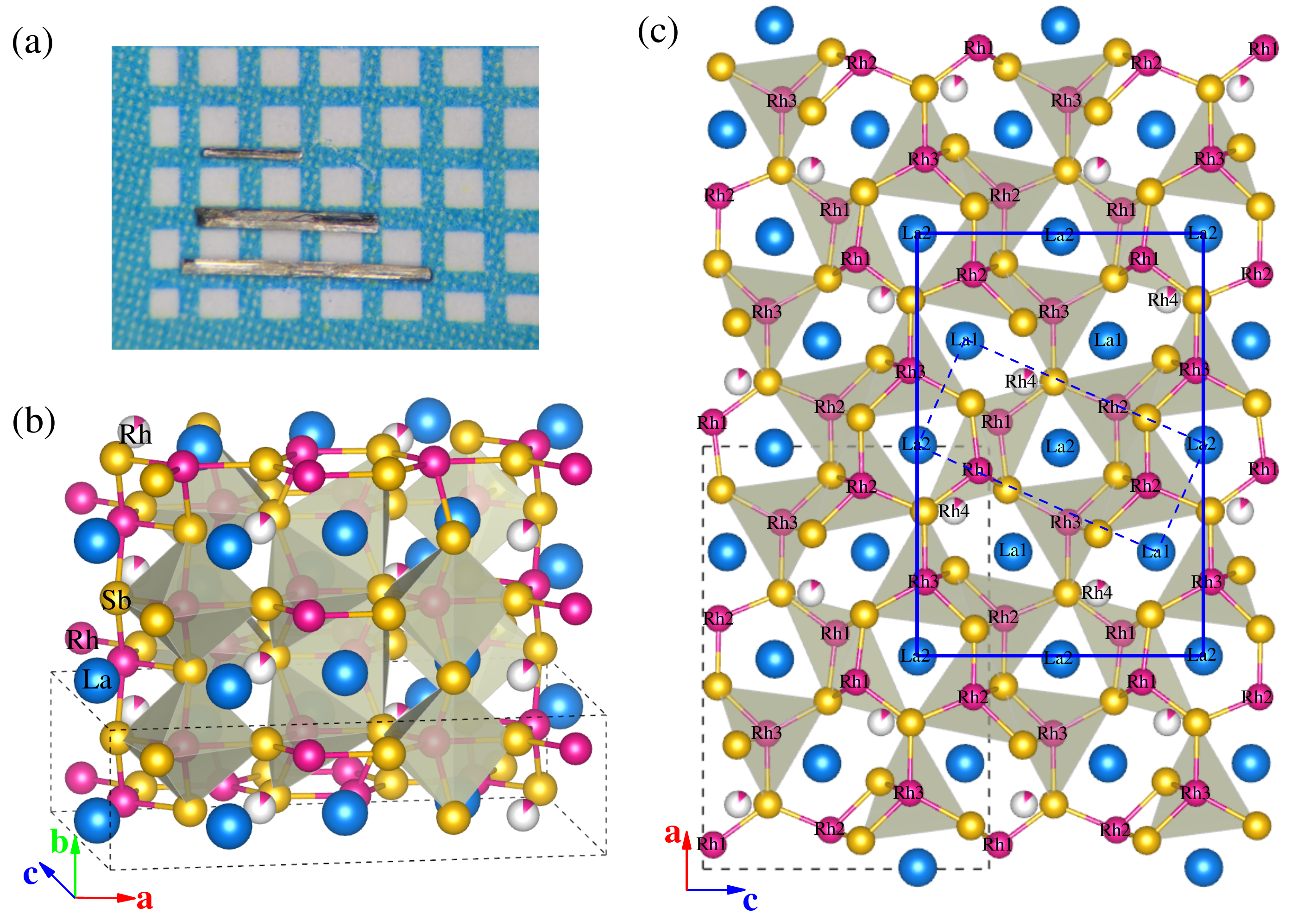}
\vspace*{10pt}
\caption{(a) A photograph of La$_2$Rh$_{3+\delta}$Sb$_4$ single crystals on millimeter-grid paper. The single crystals are needle-like along $\mathbf{b}$ axis. (b-c) show the crystalline structure. The partially occupied Rh4 is double-colored. The black-dotted rectangles mean the original unit cell as determined by XRD refinement. To better compare the cyrstalline structure to the STEM image, a translated unit cell is also shown by the blue-solid rectangle. The blue-dashed rectangular shows the part that mimics the distorted CaBe$_2$Ge$_2$-type structure.}
\label{Fig1}
\end{figure}

\section{Experimental}

Single crystals of La$_2$Rh$_{3+\delta}$Sb$_4$ were synthesized by a Bi-flux method. La chunk (alfa Aesar, 99.9$\%$), Rh powder (aladdin, 99.95$\%$), Sb granule (aladdin, 99.9999$\%$), and Bi granule (aladdin, 99.9999$\%$) were weighed in a molar ratio of 2:3:4:40. The mixture was transferred into an alumina crucible which was sealed in an evacuated quartz tube. All these operations were conducted in an argon-filled glove box to avoid the contamination of water and oxygen. The quartz tube was heated to 500 $^{\circ}$C in 20 hours and held for one day. Then, it was raised to 1100 $^{\circ}$C in 30 hours and sintered at this temperature for 100 hours. After that, in 2 weeks it was slowly cooled down to 500 $^{\circ}$C at which temperature the Bi-flux was removed by centrifugation. The remaining Bi flux can be dissolved by a mixture of an equal volume of hydrogen peroxide and acetic acid. The obtained samples were needle-like with typical dimensions $\sim 4\times 0.5 \times 0.3$ mm$^3$ [see Fig.~\ref{Fig1}(a)], and are stable in air and moisture. The same method is also applicable to grow the Ce$_2$Rh$_{3+\delta}$Sb$_4$ single crystals, a heavy-fermion compound whose low-temperature physical properties will be published elsewhere\cite{Cheng-Ce2Rh3Sb4}.

The chemical composition of the new compound was verified by energy dispersive x-ray spectroscopy (EDS) affiliated to a field emission scanning electron microscope (FEI Model SIRION), which gives the atomic ratio La:Rh:Sb = 22.2(9):33.2(4):44.7(6), close to the stoichiometric La$_2$Rh$_3$Sb$_4$, as shown in Supplementary Information (SI). The crystalline structure of
the compound was determined by a Rigaku XtaLAB mini II Single Crystal X-ray Diffractometer (XRD) with Mo radiation ($\lambda_{K\alpha}$ = 0.71073 \AA). Several single crystals were selected to collect diffraction data for structural refinement. We used the olex2-1.5 software of SHELX-2018 package and full-matrix least squares on the \emph{F$^2$} model to perform the refinement. The cell parameters and atomic coordinates were all determined, and the qualified CIF report was generated (SI). Scanning transmission electron microscope (STEM) images were recorded at 300 kV using a probe Cs-corrected Spectra 300 (ThermoFisher Ltd.) in both Dark-Field (DF) and High-Angle Annular Dark-Field (HAADF) modes. Cross-section TEM sample was prepared on the [0 1 0] plane using Helios 5UX 30 kV Ga Focused Ion Beam (FIB).

A Magnetic Property Measurement System (SQUID-VSM, Quantum Design) was employed to measure the DC magnetic susceptibility between 1.8-300 K in the presence of external field 1000 Oe ($\mathbf{H}\perp\mathbf{b}$). Electrical resistivity was measured in a standard four-probe method with current parallel to $\mathbf{b}$ axis, in a Physical Property Measurement System (PPMS-9, Quantum Design). The same instrument was also used for the specific heat measurements. For sub-Kelvin experiments, a dilution refrigerator insert was utilized to measure resistivity, AC susceptibility and specific heat. The AC susceptibility was measured at zero static magnetic field with AC driving field 0.5 Oe and various frequencies.

The electronic band structure and density of states (DOS) were calculated on the stoichiometric La$_2$Rh$_3$Sb$_4$ using density-functional theory (DFT) method as implemented in Vienna Abinit Simulation Package (VASP)\cite{method:vasp}. The plane-wave basis up to 400 eV and Perdew-Burke -Ernzerhoff (PBE)\cite{method:pbe} flavor of generalized gradient approximation to exchange-correlation functional were employed. The ion-valence electron interaction were approximated with projected augmented wave (PAW) method\cite{method:paw,method:pawvasp}. SOC was considered in all calculations. A $3\times12\times4$ Monkhorst-Pack K-grid for Brillouin zone integration was found to converge the total energy within 1 meV/atom. Using the experimental structure as initial guess, the crystal geometry was fully optimized until forces on each atom less than 1 meV/\AA\ and internal stress less than 1 kBar.

\section{Results and Discussion}

\subsection{Crystalline Structure}
Figure ~\ref{Fig1}(b) shows the crystal structure of La$_2$Rh$_{3+\delta}$Sb$_4$. It is of the orthorhombic Pr$_2$Ir$_3$Sb$_4$-like structure\cite{GilR-Ce234,Schafer-Ce234}, in the space group $Pnma$ (No. 62). During the structural refinement, we were aware of some redundant electron density near the site (0.3471, 0.25, 0.6223), and this was later recognized as the Rh4 site that has a partial occupation rate $\sim$0.122(4), by STEM experiments (see below). The refined lattice parameters are $a= 16.2667(5)$ \AA, $b=4.5858(2)$ \AA, $c=11.0068(4)$ \AA, $\alpha=\beta=\gamma=90^{\circ}$, and $Z=4$. The density of this material is 8.787 g/cm$^3$. More details about the structural information can be found in Table~\ref{tbl:Structure}, while the atomic coordination and anisotropic displacement parameters are listed in Tables~\ref{tb2:Atoms} and \ref{tbs1:Anisotropic Displacement}, respectively. The refinement results for $R_1$ and $w R_2$ are 0.0286 and 0.0560, respectively, confirming the validity of XRD refinement. It can be seen from the analysis that the structure we derived is essentially isomorphic with Ce$_2$Rh$_3$Sb$_4$ that was reported previously by polycrystal studies\cite{Schafer-Ce234,GilR-Ce234}. As expected, the lattice parameters $a$, $b$, and $c$ of Ce$_2$Rh$_3$Sb$_4$ are relatively smaller than in La$_2$Rh$_{3+\delta}$Sb$_4$\cite{GilR-Ce234}, owing to the lanthanide contraction effect.

\begin{table}[!ht]
\tabcolsep 0pt \caption{\label{tbl:Structure} Single-crystal XRD refinement data for La$_2$Rh$_{3+\delta}$Sb$_4$ at 290 K.}
\vspace*{-15pt}
\begin{center}
\def\temptablewidth{1\columnwidth}
{\rule{\temptablewidth}{1pt}}
\begin{tabular*}{\temptablewidth}{@{\extracolsep{\fill}}cc}
composition ~~~~~~~~~~~~~~~~~~~~~~~~~~~~~~~~~~~         &  La$_2$Rh$_{3+\delta}$Sb$_4$           \\
formula weight (g/mol) ~~~~~~~~~~~~~~~~~~~~      &  1086.16                     \\
space group ~~~~~~~~~~~~~~~~~~~~~~~~~~~~~~~~~~~         &  $Pnma$ (No. 62)             \\
$a$ (\AA) ~~~~~~~~~~~~~~~~~~~~~~~~~~~~~~~~~~~~~~~~~~~           &  16.2667(5)     \\
$b$ (\AA) ~~~~~~~~~~~~~~~~~~~~~~~~~~~~~~~~~~~~~~~~~~~           &  4.5858(2)      \\
$c$ (\AA)  ~~~~~~~~~~~~~~~~~~~~~~~~~~~~~~~~~~~~~~~~~~~          &  11.0068(4)   \\
$V$ (\AA$^{3}$) ~~~~~~~~~~~~~~~~~~~~~~~~~~~~~~~~~~~~~~~~~~     &  821.06(5)    \\
$Z$  ~~~~~~~~~~~~~~~~~~~~~~~~~~~~~~~~~~~~~~~~~~~~~~~~~ &  4               \\
$\rho$ (g/cm$^{3}$) ~~~~~~~~~~~~~~~~~~~~~~~~~~~~~~~~~~~~~~ &  8.787                     \\
$2\theta$ range ~~~~~~~~~~~~~~~~~~~~~~~~~~~~~~~~~~~~~~~~     &  4.468-51.998$^{\circ}$         \\
no. of reflections, $R_{int}$ ~~~~~~~~~~~~~~~~~~~~~~  &   3112, 0.0307                           \\
no. of independent reflections ~~~~~~~~~~~~ &   916                                    \\
no. of parameters ~~~~~~~~~~~~~~~~~~~~~~~~~~~ &   62                  \\
$R_1^{\dag}$, $w R_2^{\ddag}$ [$I>2\delta(I)$] ~~~~~~~~~~~~~~~~~~~~~~~~~~~&   0.0250, 0.0545    \\
$R_1$, $w R_2$ (all data) ~~~~~~~~~~~~~~~~~~~~~~~~~~~ &   0.0286, 0.0560          \\
goodness of fit on $F^2$ ~~~~~~~~~~~~~~~~~~~~~~~ & 1.135                                           \\
largest diffraction peak and hole (e/\AA$^{3}$)   &   1.92 and $-$1.90       \\
\end{tabular*}
{\rule{\temptablewidth}{1pt}}
\end{center}
$^{\dag}$ $R_1=\sum||F_{obs}|-|F_{cal}||/\sum|F_{obs}|$.~~~~~~~~~~~~~~~~~~~~~~~~~~~~~~~~~~~~~~~~~~~~~~~~~~~~~~~~~~~~~~~~~~~~~~~~~~~~~~~~  \\
$^{\ddag}$ $w R_2=[\sum w(F_{obs}^2-F_{cal}^2)^2/\sum w (F_{obs}^2)^2]^{1/2}$, $w=1/[\sigma^2F_{obs}^2+(a\cdot P)^2+b\cdot P]$, where  $P=[\mathrm{max}(F_{obs}^2)+2F_{cal}^2]/3$.~~~~~~~~~~~~~~~~~~~~~~~~~~~~~~~~~~~~~~~~~~~~~~~~~~~~~~~~~~~~~~~~~~~~~~~~~~~~~  \\
\end{table}

\begin{figure}[!ht]
\centering
\vspace*{0pt}
\hspace*{-20pt}
\includegraphics[width=17cm]{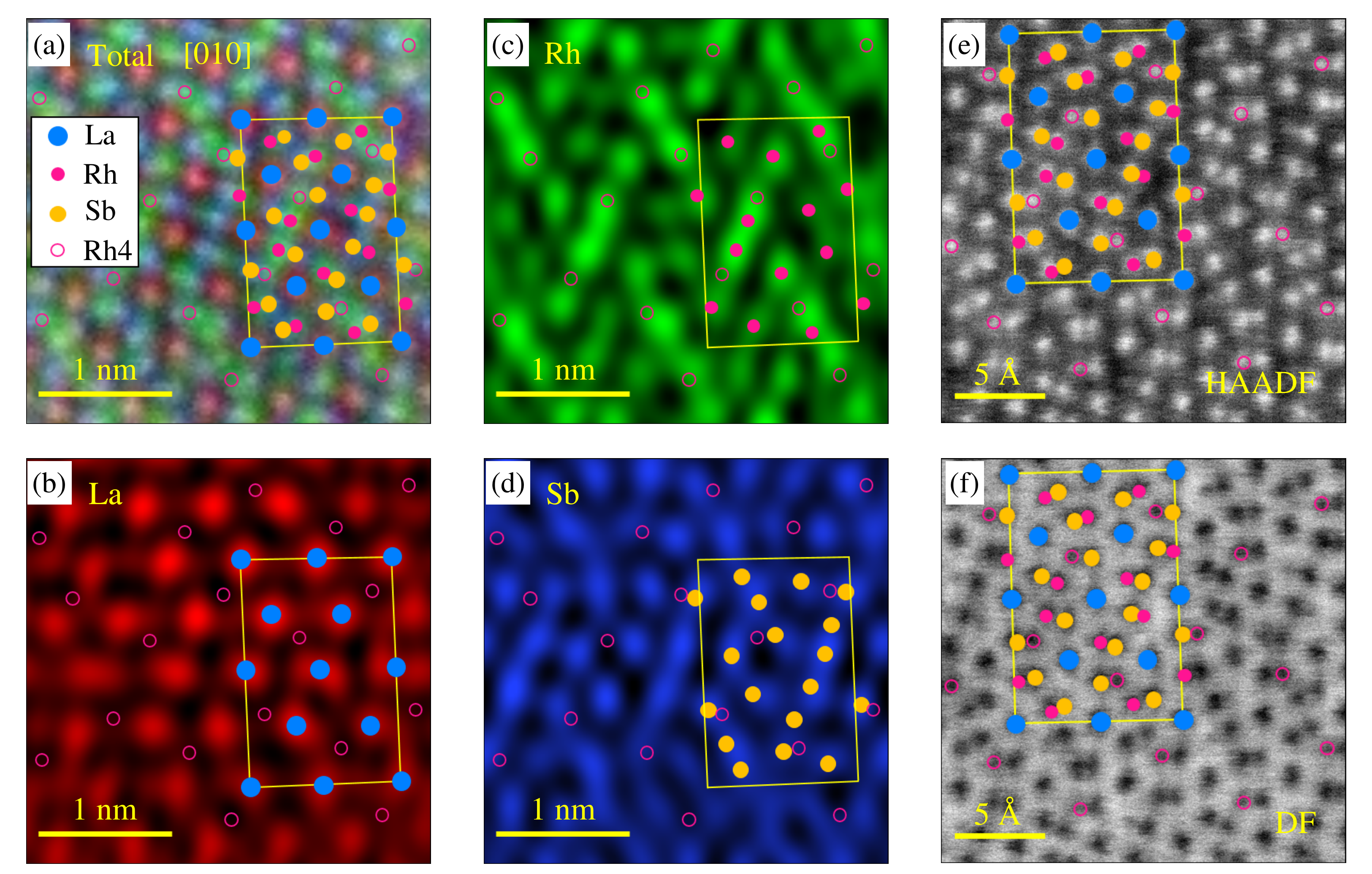}
\vspace*{10pt}
\caption{STEM images of La$_2$Rh$_{3+\delta}$Sb$_4$ on [0 1 0] plane. STEM-EDS elemental maps respectively for (a) Total, (b) La, (c) Rh, and (d) Sb. The atomic coordinates in a unit cell of La$_2$Rh$_{3+\delta}$Sb$_4$ are projected to the images. The filled blue, pink and green circles are La, Rh(1-3) and Sb atoms, respectively, while the open pink circles represent the occupied Rh4 sites. (e) HAADF image. (f) DF image.}
\label{Fig2}
\end{figure}

To clarify more details about the crystalline phase, we performed STEM measurements, as shown in Fig.~\ref{Fig2}. The high spatial resolution enables us to view individual atoms in the [0 1 0] plane, and the coordinates of the atoms are in good agreement with XRD refinement. We then project the Rh4 coordinates onto the STEM-EDS elemental maps for La, Rh, and Sb [Fig.~\ref{Fig2}(b), (c) and (d) respectively], and noticed that they fit only to the Rh image satisfactorily well, cf the open pink circles in Fig.~\ref{Fig2}(c), confirming the occupation of Rh atoms at these sites. In general, the contrast at Rh4 sites is relatively lower [Fig.~\ref{Fig2}(e-f)], because of the partial occupation. Note that the STEM sample is tens of nanometers in thickness which contains an order of magnitude 10-100 layers. Further studies based on first-principles calculations suggest that Rh4 likely serves as charge dopant, seeing below.

\begin{table}[!ht]
\tabcolsep 0pt \caption{\label{tb2:Atoms} Structural parameters and equivalent isotropic displacement parameters $U_{eq}$ of La$_2$Rh$_{3+\delta}$Sb$_4$. $U_{eq}$ is taken as 1/3 of the trace of the orthogonalised $U_{ij}$ tensor.}
\vspace*{-15pt}
\begin{center}
\def\temptablewidth{1\columnwidth}
{\rule{\temptablewidth}{1pt}}
\begin{tabular*}{\temptablewidth}{@{\extracolsep{\fill}}ccccccc}
 Atoms   &  Wyck.  &  $x$            &     $y$     &     $z$      &     Occ.      & $U_{eq}$ (\AA$^2$) \\\hline
  La1    &  4c     &  0.24998(4)     &  0.2500     &  0.91587(7)  &    1.000      &  0.0057(2)       \\
  La2    &  4c     &  0.50341(4)     &  0.2500     &  0.74938(6)  &    1.000      &  0.0080(2)       \\
  Sb1    &  4c     &  0.43260(4)     &  0.2500     &  0.05118(7)  &    1.000      &  0.0053(2)         \\
  Sb2    &  4c     &  0.65345(4)     &  0.2500     &  0.27333(7)  &    1.000      &  0.0067(2)         \\
  Sb3    &  4c     &  0.29335(5)     &  0.7500     &  0.11840(7)  &    1.000      &  0.0082(2)         \\
  Sb4    &  4c     &  0.39786(4)     &  0.2500     &  0.43616(7)  &    1.000      &  0.0091(2)         \\
  Rh1    &  4c     &  0.55719(5)     &  0.2500     &  0.46457(8)  &    1.000      &  0.0033(2)         \\
  Rh2    &  4c     &  0.59233(5)     &  0.2500     &  0.05455(8)  &    1.000      &  0.0039(2)         \\
  Rh3    &  4c     &  0.32015(5)     &  0.2500     &  0.22321(8)  &    1.000      &  0.0055(2)         \\
  Rh4    &  4c     &  0.3471(5)      &  0.2500     &  0.6223(7)   &    0.122(4)   &  0.0163(2)        \\
\end{tabular*}
{\rule{\temptablewidth}{1pt}}
\end{center}
\end{table}

\begin{figure}[!ht]
\hspace{-20pt}
\includegraphics[width=17cm]{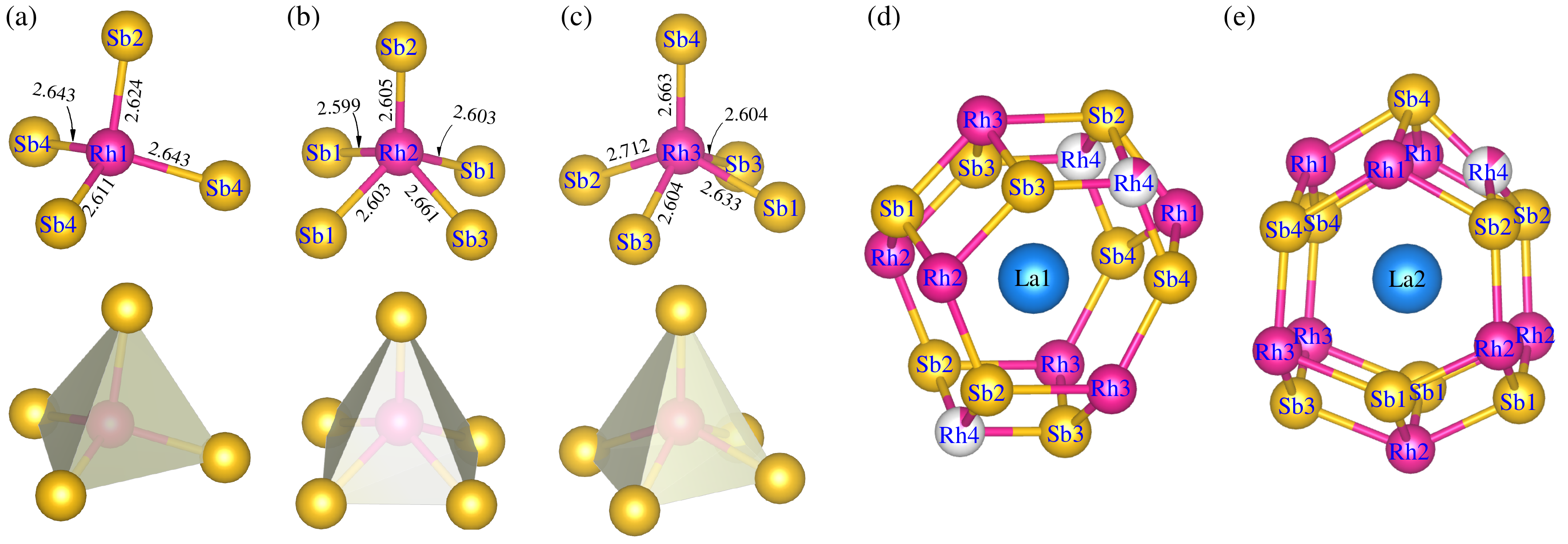}
\vspace{10pt}
\caption{Local environments in La$_2$Rh$_{3+\delta}$Sb$_4$, (a) for Rh1, (b) for Rh2, (c) for Rh3, (d) for La1, and (e) for La2. The bond lengths are in \AA. }
\label{Fig3}
\end{figure}

On the whole, the crystalline structure of La$_2$Rh$_{3+\delta}$Sb$_4$ can be viewed as stacking the Rh-Sb polyhedron networks along the $\mathbf{b}$ axis, and the La and Rh4 atoms reside inside the cavities of these networks [Fig.~\ref{Fig1}(c)]. The local environments for Rh1, Rh2 and Rh3 are depicted in Fig.~\ref{Fig3}. Rh1 is surrounded by one Sb2 and three Sb4, forming a distorted tetrahedron. Rh2 has five near-neighbor Sb atoms: three Sb1, one Sb2 and one Sb3, and they construct a hexahedron. The situation for Rh3 is similar, i.e. one Sb1, one Sb2, two Sb3 and one Sb4 also build up a hexahedron; however, here the Sb1, Sb2 and two Sb3 atoms are almost in the same plane, so the hexahedron looks rather like a pyramid. The interatomic distances for Rh-Sb bondings are shown in Fig.~\ref{Fig2}. Viewing along the $\mathbf{b}$ axis [Fig.~\ref{Fig1}(c), the area highlighted by a blue-solid rectangle], one clearly finds that two edge-sharing Rh1 tetrahedra are connected to two Rh2 and four Rh3 hexahedra by corner-sharing, of which each Rh2 hexahedron connects to one other Rh2 and Rh3 hexahedra by edge-sharing, while each Rh3 hexahedra connects to two other Rh2 hexahedra by edge-sharing and corner-sharing. In this way, a two dimensional Rh-Sb polyhedron network is weaved in the $\mathbf{ac}$-plane, and two kinds of cavities come into being: a quadrilateral type filled with La1 and Rh4, and a triangular type filled with La2. See more details in Fig.~\ref{Fig1}(c).

From the viewpoint of La, both La1 and La2 are surrounded by 18 Rh and Sb sites (including Rh4), seeing Fig.~\ref{Fig3}(d) and (e). These coordination numbers reduce to 15 and 17 respectively in the standard Pr$_2$Ir$_3$Sb$_4$ structure. Noteworthy that the local coordination for La2 mimics that of Ca in the CaBe$_2$Ge$_2$-type 122 phase, therefore, part of the structure that resembles a distorted CaBe$_2$Ge$_2$ can be recognized in La$_2$Rh$_{3+\delta}$Sb$_4$, as shown by the blue-dashed rectangular in Fig.~\ref{Fig1}(c). A clearer description can be found in Ref.~\cite{GilR-Ce234}.

\subsection{Physical Properties}

\begin{figure}[!ht]
\centering
\vspace*{-0pt}
\includegraphics[width=16cm]{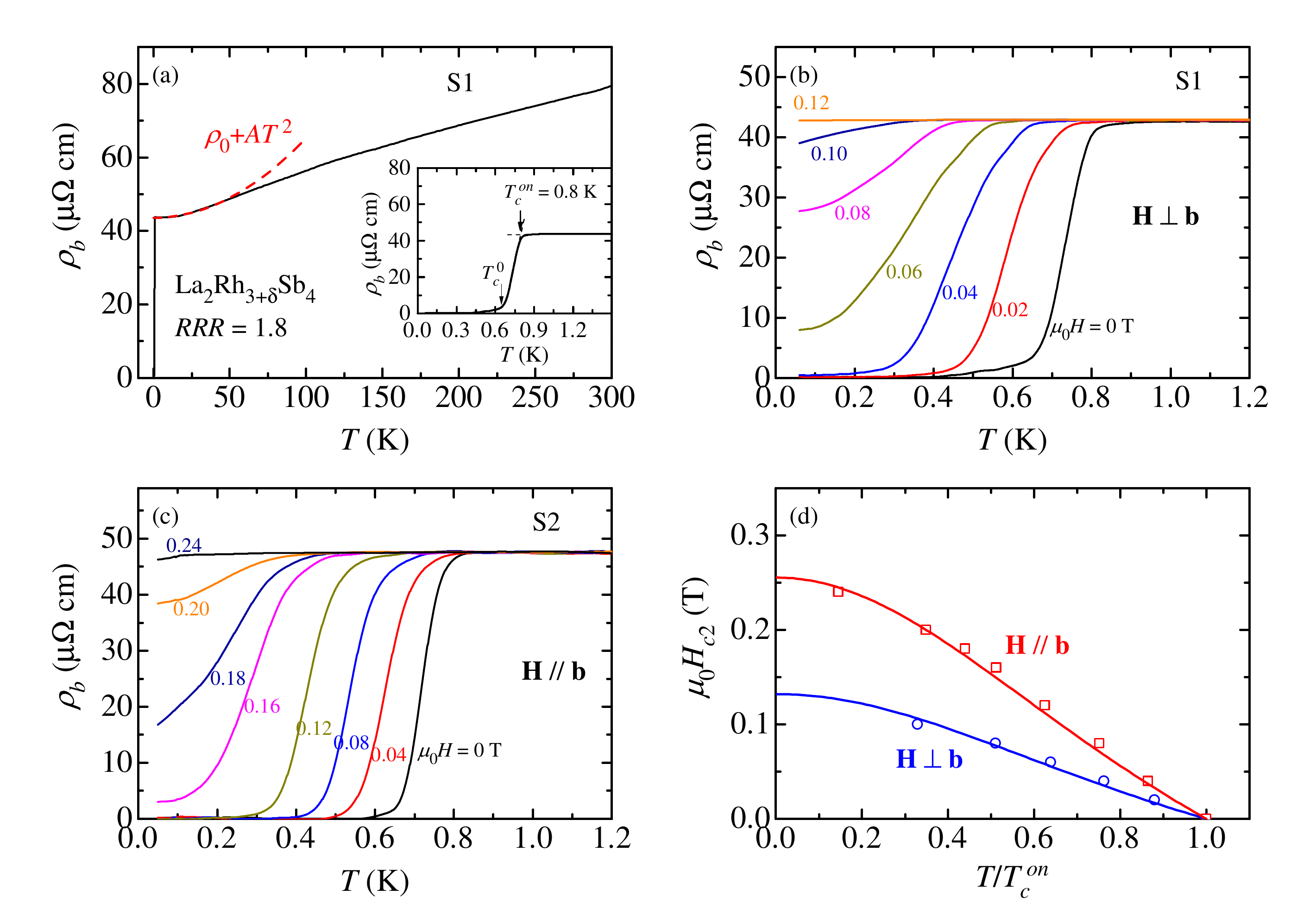}
\caption{Electrical resistivity of La$_2$Rh$_{3+\delta}$Sb$_4$. (a) $\rho_b(T)$ profile showing superconducting transition with $T^{on}_c=0.8$ K and $T_c^0=0.65$ K. (b) and (c) for $\rho_b(T)$ measured under magnetic field $\mathbf{H}\perp\mathbf{b}$ and $\mathbf{H}\parallel\mathbf{b}$, respectively. (d) Temperature dependent upper critical field $\mu_0H_{c2}(T)$; the solid lines are fits to Ginzburg-Landau law.}
\label{Fig4}
\end{figure}

Figure \ref{Fig4}(a) displays the temperature dependence of resistivity for electric current along $\mathbf{b}$. At room temperature, $\rho_b$=79.6 $\mu\Omega\cdot$cm. Upon cooling, $\rho_{b}(T)$ decreases almost linearly until for $T<50$ K a Fermi liquid behavior $\rho_b=\rho_0+AT^2$ can be well fit to the data, where the prefactor $A=2.17\times10^{-3} \mu\Omega$ cm K$^{-2}$, and $\rho_0=43.6~\mu\Omega$ cm stands for residual resistivity. This fit also yields the residual resistivity ratio $RRR\equiv\rho(300 K)/\rho_0$ = 1.8. The low $RRR$ is likely due to disorders introduced by the partially occupied Rh4. Further decreasing temperature in a dilution refrigerator, we find $\rho_b$ drops rapidly below $T_c^{on}\approx0.8$ K and essentially vanishes at $T_c^0$ = 0.65 K, indicating a SC transition. The onset of SC, $T_{c}^{on}$, is determined at which temperature $\rho_b$ drops to 99\% of $\rho_b(1 K)$, seeing the inset to Fig.~\ref{Fig4}(a).

Next, we took the resistivity measurements under magnetic field $\mathbf{H}$ perpendicular to and parallel with $\mathbf{b}$, and the results are displayed in Fig.~\ref{Fig4}(b) and (c), respectively. For $\mathbf{H}\perp\mathbf{b}$, magnetic field gradually suppresses the SC transition, and when $\mu_0H=0.12$ T, seldom resistivity drop can be seen. The upper critical field $\mu_0H_{c2}(0)$ can be obtained from the $\rho_b(H, T)$ data, by fitting the field dependent $T_c^{on}$ to the Ginzburg-Landau formula\cite{GL-1950}, $\mu_0H_{c2}(T)=\mu_0H_{c2}(0)(1-t^2)/(1+t^2)$, where $t=T/T_c^{on}$ is the reduced temperature. This fit results in $\mu_0H_{c2}^{\perp b}(0)=0.13$ T. The measurements for $\mathbf{H}\parallel\mathbf{b}$ were made on another sample whose SC transition is similar. In this field orientation, we derived $\mu_0H_{c2}^{\parallel b}(0)=0.26$ T. The anisotropy of upper critical field thus is $\sim$ 2. These $\mu_0H_{c2}$ values are far less than the Pauli paramagnetic limit $\mu_0H_P=1.84 T_c\approx1.4$ T \cite{Clogston-PauliLimit,Chandrasekhar-PauliLimit}, implying that orbital limiting is the dominant.

\begin{figure}[!ht]
\centering
\vspace*{-20pt}
\hspace*{0pt}
\includegraphics[width=9.5cm]{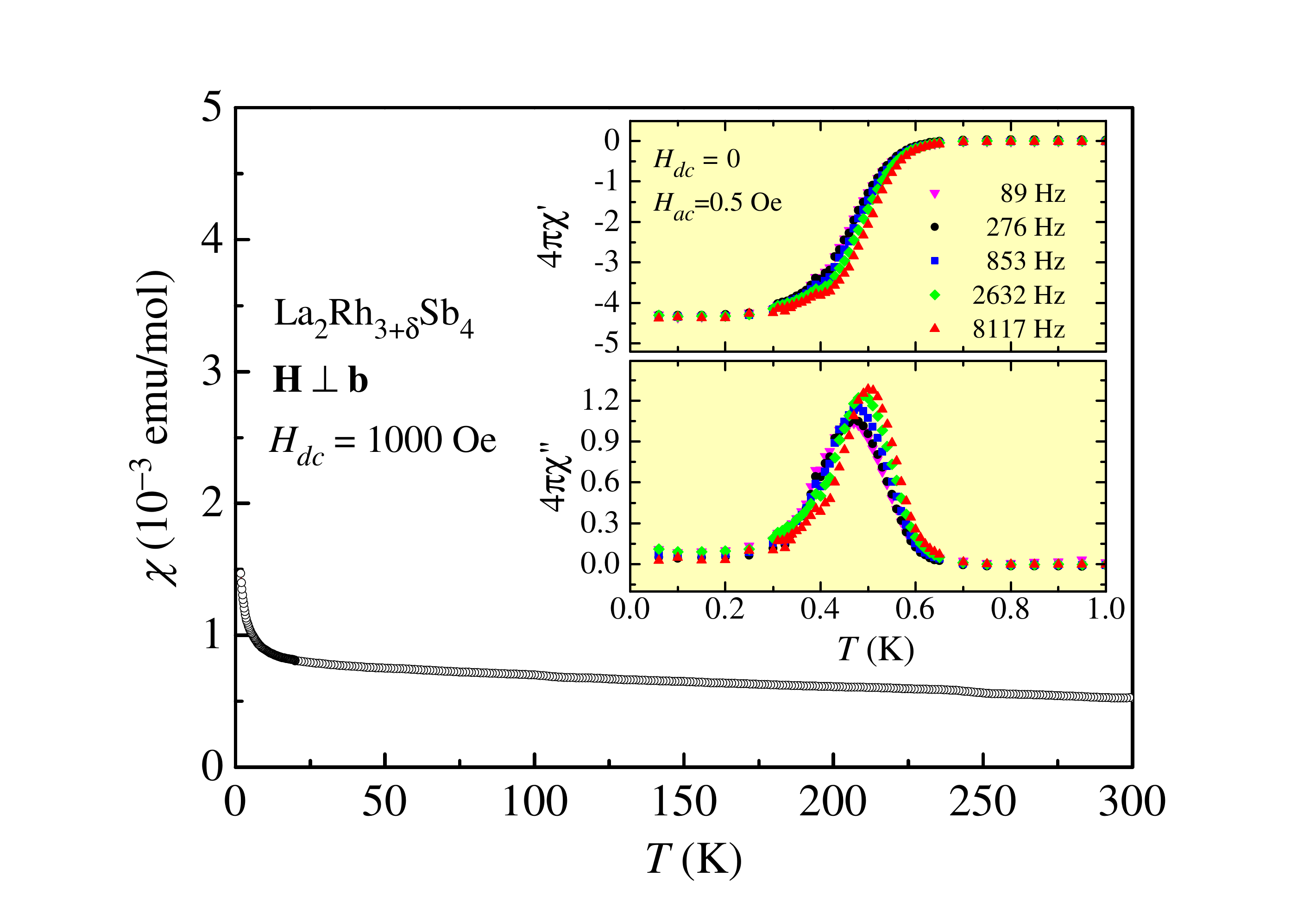}
\caption{Temperature dependence of DC magnetic susceptibility for field perpendicular to $\mathbf{b}$. The insets present real ($\chi'$) and imaginary ($\chi''$) parts of AC susceptibility as functions of $T$, measured at various frequencies.}
\label{Fig5}
\end{figure}

The SC of La$_2$Rh$_{3+\delta}$Sb$_4$ is further confirmed by magnetic shielding effect, as shown in Fig.~\ref{Fig5}. Our DC magnetic susceptibility results reveal a weak paramagnetic property above 2 K, characteristic of itinerant metals with little local moment. Note that the upturn tail at low temperature is probably due to a tiny amount of magnetic impurity. AC susceptibility was employed for the sub-Kelvin region. The real part of AC susceptibility $\chi'$ drops rapidly below 0.65 K and the crystal becomes diamagnetic. We should point out that the transition temperature determined by $\chi'(T)$ is very close to $T_c^0$ from resistivity measurements, manifesting that bulk SC is realized below 0.65 K. Far below $T_c$, the magnetic shielding fraction of SC (without taking demagnetization-factor correction) exceeds 400\% (cf upper inset to Fig.~\ref{Fig5}). Furthermore, we also notice that the SC transition is frequency dependent, \textit{i.e.}, the peak of $\chi''(T)$, the imaginary part of AC susceptibility, moves to higher temperature as the frequency increases. This is reminiscent of dissipation caused by flux motion in the mixed state of SC, and implies that La$_2$Rh$_{3+\delta}$Sb$_4$ is a type-II superconductor.

\begin{figure}[!ht]
\centering
\vspace*{-20pt}
\hspace*{0pt}
\includegraphics[width=9.5cm]{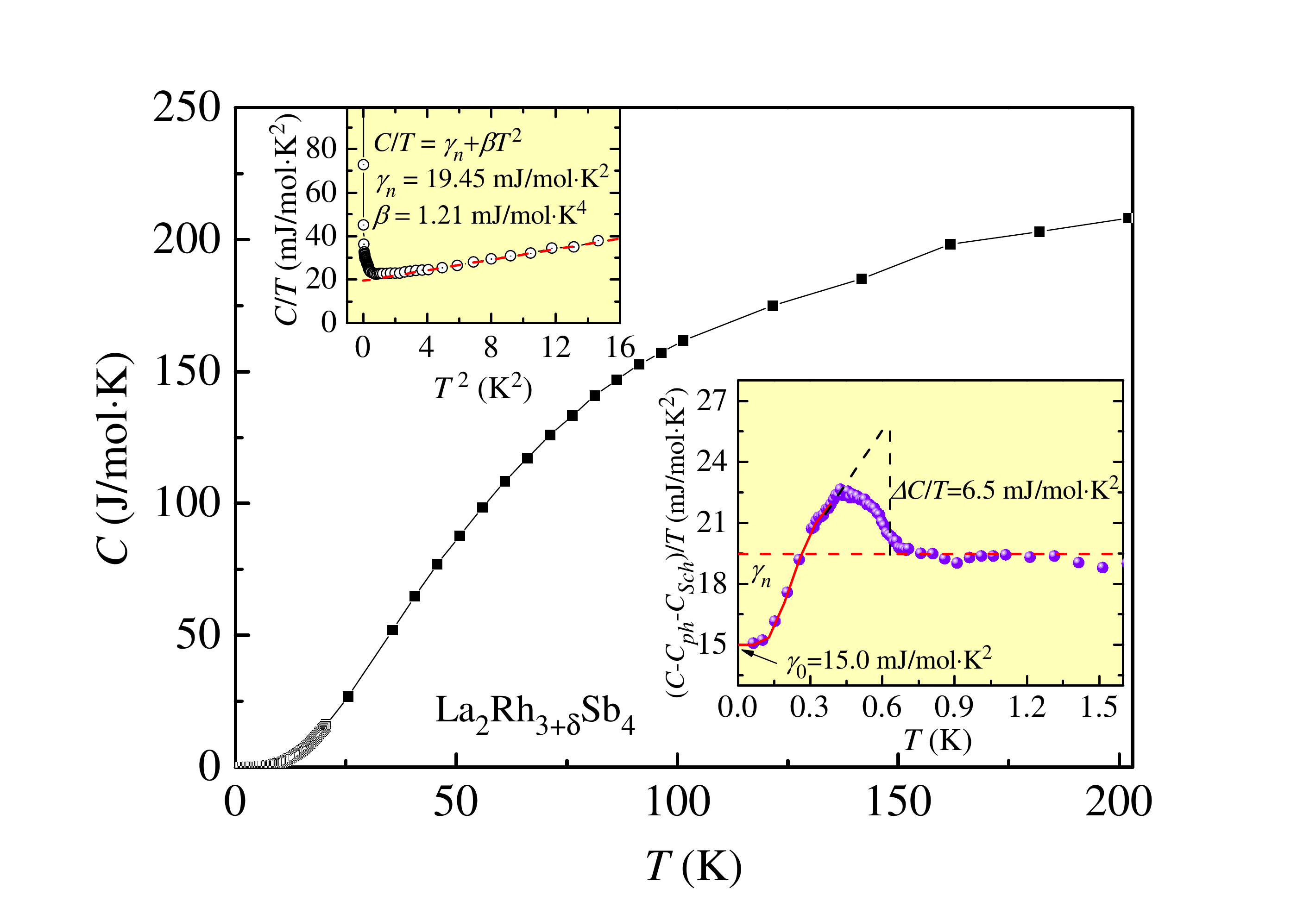}
\caption{Temperature dependent specific heat ($C$) of La$_2$Rh$_{3+\delta}$Sb$_4$. The left inset shows a plot of $C/T$ vs. $T^2$ in the low temperature range. The upturn of $C/T$ for $T\rightarrow 0$ arises from the nuclear Schottky anomaly. The bottom inset, $(C-C_{ph}-C_{Sch})/T$ as a function of $T$, displays the jump due to SC transition $\Delta C/T|_{T_c}$=6.5 mJ/mol$\cdot$K$^2$.}
\label{Fig6}
\end{figure}

To further study the SC properties of La$_2$Rh$_{3+\delta}$Sb$_4$, we turn to specific heat measurements, the results of which are summarized in Fig.~\ref{Fig6}. At 200 K, the total specific heat $C$ is about 206 J/mol$\cdot$K, close to the classic limit 3$NR$ = 224.5 J/mol$\cdot$K given by Dulong-Petit's law\cite{Ashcroft-SSP}, where $N$ = 9 is the number of atoms in a formula unit of stoichiometric La$_2$Rh$_3$Sb$_4$, and $R$ = 8.314 J/mol$\cdot$K is the ideal gas constant. The low temperature specific heat conforms to $C(T)=\gamma_n T+\beta T^3$, where the first and second terms are contributions from electrons ($C_{el}$) and phonons ($C_{ph}$). This is better expressed by the $C/T$ vs. $T^2$ plot inset to Fig.~\ref{Fig6}. This fit leads to the normal-state Sommerfeld coefficient $\gamma_n=19.45$ mJ/mol$\cdot$K$^2$, and $\beta=$ 1.21 mJ/mol$\cdot$K$^4$, and enables us to estimate the Debye temperature, $\Theta_D=(\frac{12\pi^4NR}{5\beta})^{1/3}=244$ K. Note that the rapid increase of $C/T$ for $T\rightarrow0$ shown in the upper inset of Fig.~\ref{Fig6} arises from the nuclear Schottky anomaly ($C_{Sch}$) and can be properly subtracted by fitting $C_{Sch}$ to a $1/T^2$ law\cite{Movshovich-PrInAg2}. We then obtain $C_{el}$ by subtracting both $C_{ph}$ and $C_{Sch}$ from the total $C(T)$, as shown in the lower inset to Fig.~\ref{Fig6}. The anomaly in specific heat due to the SC transition now can be clearly seen near 0.65 K, and the jump of specific heat at $T_c$ is estimated $\Delta C/T|_{T_c}=6.5$ mJ/mol$\cdot$K$^2$. Below $T_c$, the electronic specific heat consists of two parts, viz. $C_{el}(T)=\gamma_0 T+C_{sc}(T)$, where $\gamma_0$ is the residual Sommerfeld coefficient, and $C_{sc}$ describes the superconducting part of specific heat. We find that $C_{sc}$ agrees well to the $s$-wave model\cite{Taylor-HeatSC}, viz. $C_{sc}/T\propto T^{-5/2}\exp{(-\Delta_0/T)}$, in which $\Delta_0$ characterizes the superconducting gap. This fit gives rise to $\gamma_0=15.0$ mJ/mol$\cdot$K$^2$ and $\Delta_0=$ 1.1 K. This manifests that only $\sim 23\%$ [$=(\gamma_n-\gamma_0)/\gamma_n$] of the density of states at Fermi energy are gapped out due to the SC transition. One possible reason for this is the multi-band effect, (see Fig.~\ref{Fig7}), as Cooper pairs probably form only in some specific bands. Similar observation has been found in other multi-band superconductors, like PrNiAsO\cite{Li-LnNiAsO}. An alternative possibility could be due to some non-superconducting fraction. It is difficult to rule out this possibility only from the present data. Microscopic and local measurements like nuclear magnetic resonance (NMR), scanning tunneling microscope (STM), and angle-resolved photoemission spectroscopy (ARPES) may be useful to clarify this issue. The analysis of specific heat further leads to two important ratios, $\frac{\Delta C}{(\gamma_n-\gamma_0)T_c}=1.46$ and $\Delta_0/T_c=$ 1.69. These values respectively are close to 1.43 and 1.76 as predicted by the BCS's theory\cite{Bardeen1957,Bardeen1957a}. La$_2$Rh$_{3+\delta}$Sb$_4$, therefore, is probably an $s$-wave superconductor.

\subsection{Electronic Structure}
The fully optimized lattice constants of \textit{pristine} La$_2$Rh$_3$Sb$_4$ in DFT calculations are $a=16.3862$ \AA, $b=4.6195$ \AA, and $c=11.0796$ \AA. Compared to the experimental results, they are slightly increased, but all are within 3\% error limit of the PBE functional, suggesting the validity of our calculations. Moreover, all relaxed internal coordinates are also accurate within 1\% of the experimental values. Usually, such pronounced accuracy in plain DFT calculations is a hint of weak correlation of the compound.

\begin{figure}[!ht]
\centering
\vspace*{-120pt}
\hspace*{-10pt}
\includegraphics[width=16.5cm]{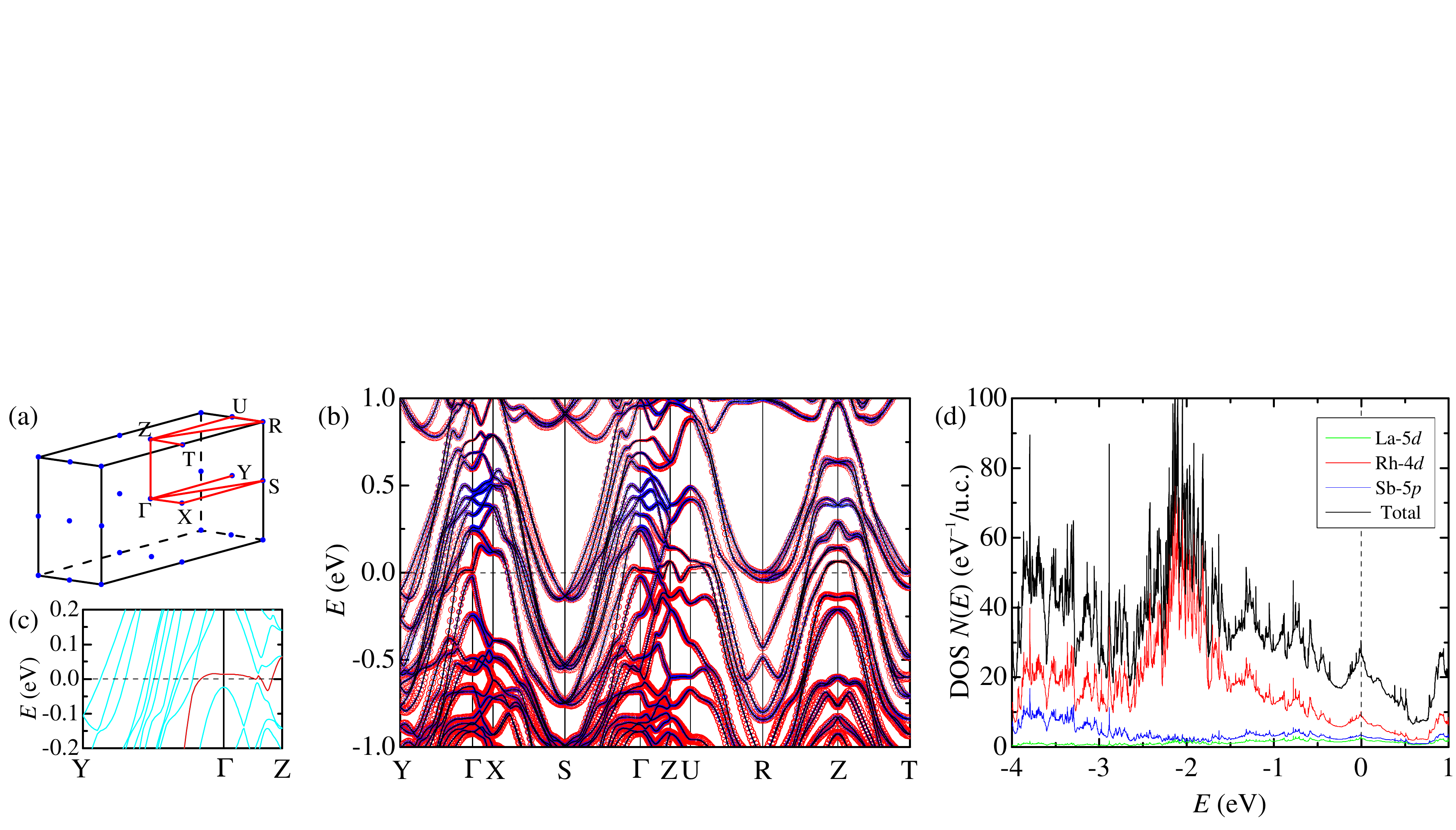}
\caption{First-principles electronic structure of La$_2$Rh$_3$Sb$_4$. (a) Sketch of Brillouin zone. (b) Band structure where the size of the red and blue circles represent the weight of Rh-4$d$ and Sb-5$p$ orbitals, respectively. (c) Enlarged band structure along Y-$\Gamma$-Z paths to show the semi-flat feature near $\Gamma$. (d) Density of states. For pristine La$_2$Rh$_3$Sb$_4$, Fermi level likely sits at a van-Hove singularity where $N(E)$ peaks.}
\label{Fig7}
\end{figure}

We show the calculated band structure and DOS in Fig.~\ref{Fig7}. The electronic DOS near the Fermi level are dominated by Rh-$4d$ and Sb-$5p$ orbitals. The La-5$d$ orbitals have only negligible contribution. The La-4$f$ states are approximately 2 eV above the Fermi level and are irrelevant. Since the crystal structure contains inversion center, the spin degeneracy is preserved with SOC considered. As the crystal structure of the compound contains 4 formula units of La$_2$Rh$_3$Sb$_4$, its band structure is quite complicated too. There are as many as 12 doubly degenerate bands crossing the Fermi level, thus giving rise to 12 doubly degenerate Fermi surface sheets. Around $\Gamma$, semi-flat bands appear along $\Gamma$-Y and $\Gamma$-Z, seeing Fig.~\ref{Fig7}(c). As a result, a small DOS peak appear at the Fermi level in the pristine compound. These semi-flat bands may be related to reduced dimensionality in certain directions, as two of these bands do not cross the Fermi level along U-R and Z-T. Therefore, the DOS peak at $E_F$ in the pristine La$_2$Rh$_3$Sb$_4$ may be interpreted as van-Hove singularity like, and yields a large $N(E_F)=7.31$ states/(f.u.$\cdot$eV), or equivalently $\gamma_{\mathrm{calc}}=$ 17.24 mJ/mol$\cdot$K$^2$. Compared to experimental value assuming $\gamma_n=\gamma_{\mathrm{calc}}(1+\lambda_{\mathrm{elph}})$, we obtain a small electron-phonon coupling constant $\lambda_{\mathrm{elph}}=$ 0.12, which cannot account for the observed $T_c$. Alternatively, if we take the inter-site Rh4 impurities as charge dopants, we can estimate the amount of electron doping to be about 1 electron/unit cell\cite{note1}. Employing the rigid band approximation, the DOS at the Fermi level is found to be $N(E_F)=$ 6.0 states/(f.u.$\cdot$eV), or equivalently $\gamma_{\mathrm{calc}}=$ 14.14 mJ/mol$\cdot$K$^2$ for the doped system. Thus, the electron-phonon coupling constant of the La$_2$Rh$_{3+\delta}$Sb$_4$ can be estimated to be $\lambda_{\mathrm{elph}}$ = 0.375. $T_c$ can be estimated according to the McMillan equation in the weak electron-phonon coupling regime\cite{McMillan-ElPhSC},
\begin{equation}
T_c=\frac{\Theta_D}{1.45}\exp[-\frac{1.04(1+\lambda_{\mathrm{elph}})}{\lambda_{\mathrm{elph}}-\mu^*(1+0.62\lambda_{\mathrm{elph}})}],
\label{Eq1}
\end{equation}
where $\mu^*$ is Coulomb pseudopotential. This yields $T_c$ = 0.6 K, assuming a commonly used value $\mu^*=$ 0.10 \cite{Mu-La1111ElPh}. The nice agreement signals BCS weak-coupling scenario and conventional electron-phonon coupled pairing mechanism.\\

\subsection{Discussion Added}
Some additional questions may have been invoked by this work. (1) It is still unknown how the pristine compound La$_2$Rh$_3$Sb$_4$ behave at low temperature, and whether and how superconductivity depends on the the Rh4 concentration $\delta$. Since stoichiometric La$_2$Rh$_3$Sb$_4$ sits at the van Hove singularity, and the $RRR$ may also be enhanced with less disorder, one might expect a higher $T_c$, assuming the electron-phonon coupling constant retains. However, this really depends. (2) According to what we have known from La$_2$Rh$_{3+\delta}$Sb$_4$ and Ce$_2$Rh$_{3+\delta}$Sb$_4$\cite{Cheng-Ce2Rh3Sb4}, the values of $\delta$ both are close to 1/8. It will be interesting to investigate whether these Rh dopants form a specific order (\textit{e.g.} the famous ``1/8 anomaly" in the high $T_c$ cuprate La$_{1.875}$Ba$_{0.125}$CuO$_4$\cite{Moodenbaugh-La214_Ba}). In recent years, many antimonide superconductors have been found to be in close relation with charge order, \textit{e.g.} CsV$_3$Sb$_5$\cite{Wang-CsV3Sb5}, IrSb\cite{Qi-IrSb,Ying-IrSb}. This remains an open issue in La$_2$Rh$_{3+\delta}$Sb$_4$. (3) Our work may also imply that the broad $Ln_2$$Tm_{3+\delta}$Sb$_4$  family may host new quantum material bases where new superconductors, quantum magnetism (note that geometrical frustration probably exists due to the triangular $Ln$ sublattice), Kondo effect and other electronic correlation effects could be found. To address these issues, more investigations are needed in the future.

\section{Conclusion}

We successfully synthesized a new ternary rhodium-antimonide La$_2$Rh$_{3+\delta}$Sb$_4$ by a Bi-flux method. The compound crystallizes in the orthorhombic Pr$_2$Ir$_3$Sb$_4$-like structure with redundant Rh acting as charge dopant. The refined crystalline structure has been determined by single-crystal X-ray diffraction and scanning transmission electron microscope. First-principles band structure calculations point out that it is a multi-band metal with a van-Hove singularity like feature close to the Fermi level. Ultra-low temperature resistivity, magnetic susceptibility, and specific heat measurements unveil it as a candidate of weak-coupling $s$-wave type-II superconductor.\\

\textbf{Supporting Information Availalbe}\\
Figure S1 shows the energy dispersive X-ray spectrum on a single crystal of La$_2$Rh$_{3+\delta}$Sb$_4$. Table S1 displays the anisotropic displacement parameters. The CIF file of La$_2$Rh$_{3+\delta}$Sb$_4$ and a report file are also provided.

\section*{Acknowledgments}

This work was supported by the open research fund of Songshan Lake Materials Laboratory (2022SLABFN27), NSF of China (Grants Nos. 12004270, 51861135104, 11574097, 11874137, 12274364, 12204298), the Fundamental Research Funds for the Central Universities of China (2019kfyXMBZ071), Guangdong Basic and Applied Basic Research Foundation (2019A1515110517), and the Pioneer and Leading Goose R\&D Program of Zhejiang (2022SDXHDX0005).

\section*{References}

\begin{thebibliography}{10}
\bibitem{note1}
  The Rh4 atom is assumed to be divalent in the calculation. Taking $Z$=4 and dopant concentration $\delta$=1/8 for simplicity, we get 1 electron/u.c. Therefore, the Fermi level is assumed to change by $\Delta E$=39 meV given by $\int_0^{\Delta E}N(E)dE=1$. $N(\Delta E)$=24 eV$^{-1}/$u.c.
\end{thebibliography}

\providecommand{\newblock}{}

\newpage
\thispagestyle{empty}

\setcounter{table}{0}
\setcounter{figure}{0}
\setcounter{equation}{0}
\setcounter{section}{0}
\renewcommand{\thefigure}{S\arabic{figure}}
\renewcommand{\thetable}{S\arabic{table}}
\renewcommand{\theequation}{S\arabic{equation}}

\begin{center}
{\it\textbf{Supplementary information: }} \\
\textbf{La$_2$Rh$_{3+\delta}$Sb$_4$: A new ternary superconducting rhodium-antimonide}\\

\end{center}

\begin{center}
Kangqiao Cheng$^{1\dag}$, Wei Xie$^{1\dag}$, Shuo Zou$^1$, Huanpeng Bu$^2$, Jin-Ke Bao$^{3}$, Zengwei Zhu$^{1}$, Hanjie Guo$^{2*}$, Chao Cao$^{4*}$, and Yongkang Luo$^{1*}$\\

\end{center}

\vspace{-140pt}
\begin{figure}[!ht]
\includegraphics[width=16cm]{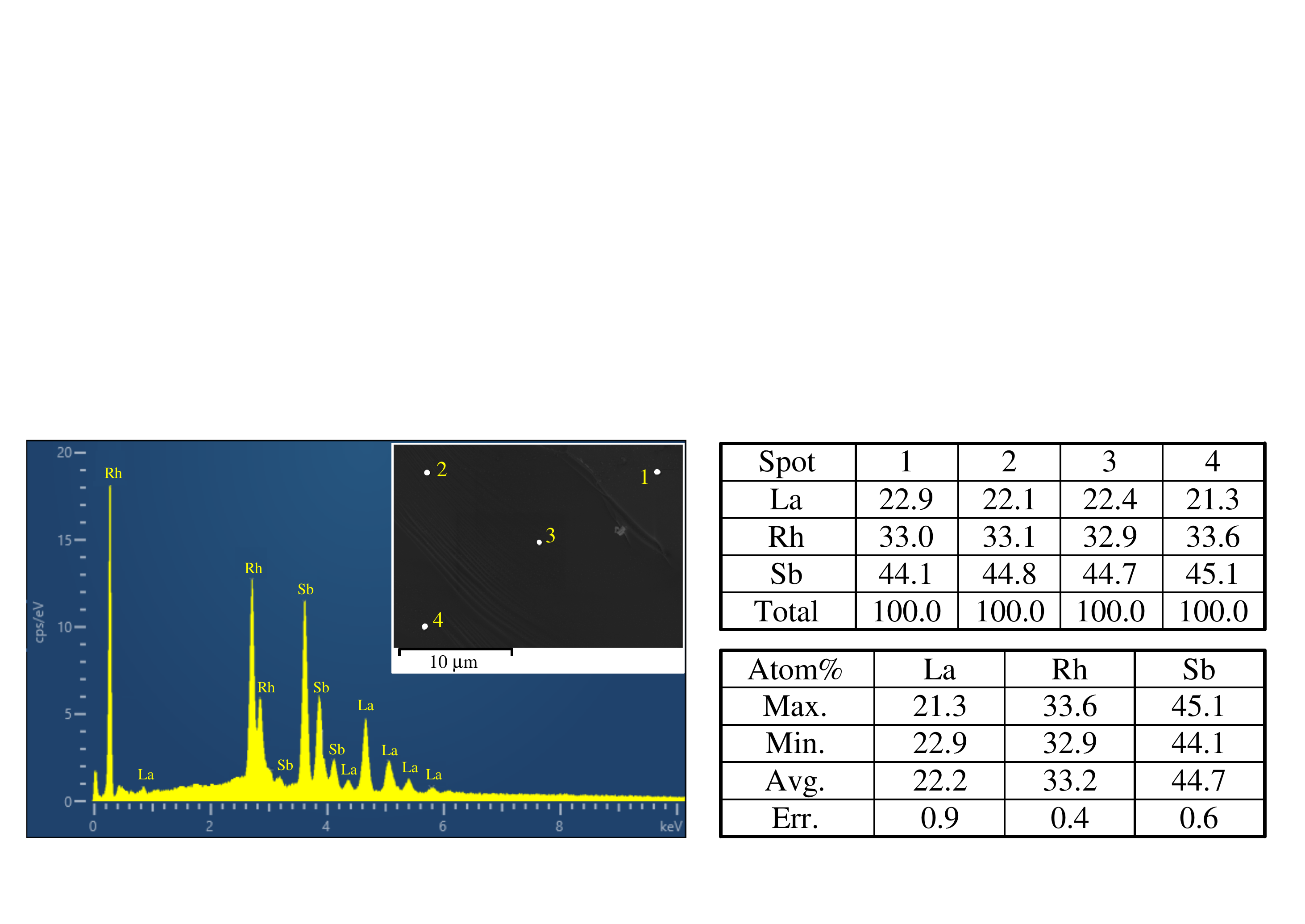}
\label{FigS1}
\vspace{-20pt}
\caption{Energy dispersive X-ray spectrum (EDS) results of a selected La$_2$Rh$_{3+\delta}$Sb$_4$ single crystal. The tables display the chemical compositions.}
\end{figure}

\begin{table}[!ht]
\tabcolsep 0pt \caption{\label{tbs1:Anisotropic Displacement}Anisotropic displacement parameters (\AA$^2$$\times 10^3$) for La$_2$Rh$_{3+\delta}$Sb$_4$. The Anisotropic displacement factor exponent takes the form: $-2\pi^2[h^2 a^{\ast 2}U_{11}+2hka^{\ast} b^{\ast} U_{12}+...]$.}
\vspace*{-15pt}
\begin{center}
\def\temptablewidth{1\columnwidth}
{\rule{\temptablewidth}{1pt}}
\begin{tabular*}{\temptablewidth}{@{\extracolsep{\fill}}ccccccc}
  Atoms &   $U_{11}$   &   $U_{22}$  &  $U_{33}$   &  $U_{23}$   &  $U_{13}$   &   $U_{12}$   \\  \hline
  La1   &    5.1(3)	   &  7.4(4)	 & 4.6(4)      &    0        &  -1.3(2)    &     0        \\
  La2   &    10.0(3)   &  8.6(4)     &	5.4(4)	   &    0        &  1.4(3)     &     0        \\
  Sb1   &    3.7(4)    &   7.5(5)	 & 4.8(4)      &    0        &  0.0(3)     &     0        \\
  Sb2   &    3.4(3)	   & 14.3(5)	 & 2.4(4)	   &    0        &  0.2(3)	   &     0        \\
  Sb3   &    12.7(4)   &	6.2(4)   &	5.7(4)	   &    0        &  4.8(3)	   &     0        \\
  Sb4   &    6.6(4)    &	7.0(5)   &	13.7(4)    &    0        &  -5.1(3)    &     0        \\
  Rh1   &    2.9(4)    &	5.4(5)   & 1.6(4)	   &    0        &  -0.3(3)    &     0        \\
  Rh2   &    3.4(4)    &	5.8(5)	 & 2.6(4)      &    0        &  -0.2(3)    &     0        \\
  Rh3   &    6.3(4)    &	7.2(5)	 & 3.0(5)	   &    0        &  1.2(3)     &     0        \\
  Rh4   &    12(4)     &    32(5)	 & 5(4)        &    0        &  3(3)       &     0        \\
\end{tabular*}
{\rule{\temptablewidth}{1pt}}
\end{center}
\end{table}

\end{document}